\documentclass[aps,pra,reprint,showpacs]{revtex4-1}
\usepackage{graphicx, amsmath}

\begin{document}
\title{Holographic power-law traps for the efficient production of Bose-Einstein condensates}
\date{\today}
\author{Graham D. Bruce}
%\altaffiliation[Present address:]{Department of Physics, SUPA, University of Strathclyde, Glasgow G4 0NG, UK}
\author{Sarah L. Bromley}
\author{Giuseppe Smirne}
\author{Lara \surname{Torralbo-Campo}}
\author{Donatella Cassettari}
\email{dc43@st-and.ac.uk}
\affiliation{Scottish Universities Physics Alliance, School of Physics and Astronomy, University of St Andrews, North Haugh, St Andrews, Fife KY16 9SS, UK}


\begin{abstract}
	We use a phase-only spatial light modulator to generate light distributions in which the intensity decays as a power law from a central maximum, with order ranging from $2$ (parabolic) to $0.5$. We suggest that a sequence of these can be used as a time-dependent optical dipole trap for all-optical production of Bose-Einstein condensates in two stages: efficient evaporative cooling in a trap with adjustable strength and depth, followed by an adiabatic transformation of the trap order to cross the BEC transition in a reversible way. Realistic experimental parameters are used to verify the capability of this approach in producing larger Bose-Einstein condensates than by evaporative cooling alone.
\end{abstract}

\maketitle

\section{Introduction}

In recent years, several techniques have been developed with the aim of producing arbitrary, time-dependent optical potentials for ultracold atoms.  A fast-scanning acousto-optic deflector can ``paint'' optical potentials with various geometries \cite{Henderson_09,Zimmermann_11,Schnelle_08}. Alternatively, the diffraction of a laser beam by a spatial light modulator (SLM) can produce both discrete \cite{McGloin_03,Bergamini_04,Boyer_06, He_09,Kruse_10} and continuous \cite{Rhodes_06,Fatemi_07, Pasienski_08,Bruce_11} holographic traps. Arbitrary trapping potentials are particularly appealing in view of quantum information processing \cite{Beugnon_07,Porto_03} and quantum simulation with neutral atoms \cite{Lewenstein_07}, the study of superfluid flow in engineered waveguides \cite{Ramanathan_11}, and for matter-wave interferometry. 

To date, whenever light patterns produced with these techniques are used to trap a quantum gas, the experimental sequence requires an additional trap (optical or magnetic), in which laser-cooled atoms are evaporatively cooled to quantum degeneracy before being transferred into the optical potential of choice. This procedure would be significantly simplified if the laser-cooled atoms were trapped directly in a holographic trap and brought to degeneracy via a dynamic transformation of the potential.  The required apparatus is also very compact: all that is needed to achieve quantum degeneracy and any subsequent manipulation of the quantum gas in the holographic trap are two crossed laser beams each modulated by an SLM.  In this paper we present power-law holographic traps suitable for this purpose, and we devise a sequence comprising a first stage of evaporative cooling, followed by a second stage in which the BEC transition is crossed reversibly by means of a change in trap order, similar to the one described in \cite{Pinkse_97}.

The paper is organised as follows: in Sec. \ref{sect:generate} we discuss the method by which one can program an SLM to generate power-law intensity distributions, and show the actual light patterns produced by the physical device. To our knowledge, this is the first time that a linear (order 1) power-law optical trap, analogous to a quadrupolar magnetic trapping potential, is suggested. We demonstrate the feasibility of using a sequence of such light patterns as a time-dependent trapping potential, by verifying that the light level does not fluctuate as the SLM is refreshed. In Sec. \ref{sect:adiabaticity} we lay out the adiabaticity conditions, with respect to the atomic motion in the trap, which any sequence must satisfy. Sec. \ref{sect:evap} details a fast and efficient evaporative cooling by trap deformation, while in Sec.~\ref{sect:transform} we analyse the adiabatic transformation of the trap in the vicinity of the BEC transition. We find that a large gain in phase-space density can be achieved during this last stage, with no significant atom loss. This contributes to the overall efficiency of the scheme in terms of the fraction of atoms left at degeneracy.

\section{Generating Holographic Traps} \label{sect:generate}

Using an iterative Fourier transform algorithm and an SLM, we can create arbitrary light patterns, including power-law intensity distributions $I\left(r\right)$ of order $\alpha$:

\begin{equation}
I\left(r\right)  = \label{eqn:I}
\begin{cases}
 I\left(0\right)\left[1-\left(\frac{r}{r_{0}}\right)^\alpha\right] \ \ \  & r\leq r_{0}, \\
0 & r>r_{0}. 
\end{cases}
\end{equation}

Atoms in this light field will, via the AC-Stark effect, experience a conservative trapping potential for $r\leq r_{0}$: 

\begin{equation}
U\left(r\right) = A \left(\frac{r}{r_{0}}\right)^\alpha,  \label{eqn:trappingpot}
\end{equation}
	
\noindent where $r_{0}$ is the trap radius and $A=3 \pi c^2 \Gamma I(0)/2 \omega_{0}^3 |\delta|$ the trap depth \cite{Grimm_00}. Here we assume light detuned by $\delta$ from the atomic transition frequency $\omega_0$. $\Gamma$ is the natural linewidth of the atomic transition and $c$ is the speed of light.

\begin{figure}%[ht!]
\includegraphics[width=0.4\textwidth]{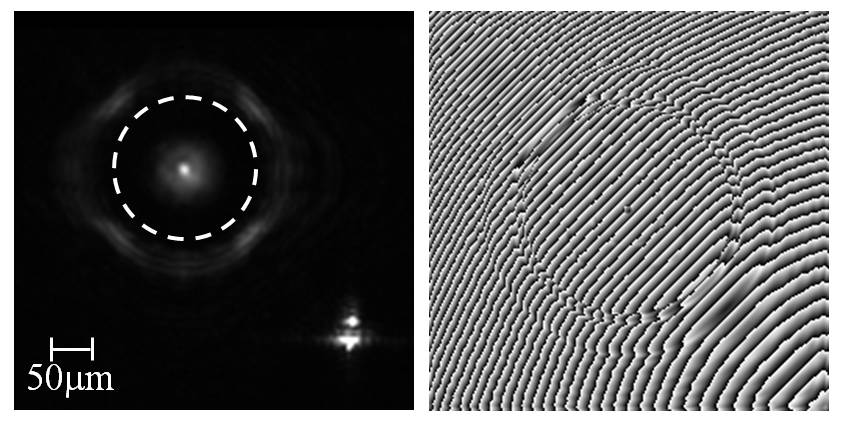}
\caption{\label{fig:Output}Left: Far-field intensity distribution of a phase-modulated laser beam. The plane is divided into a signal region (inside the circle) containing a power-law distribution of order 2, and a noise region where the intensity is unconstrained. The bright spot in the bottom right is undiffracted light. This intensity distribution is generated holographically by illuminating a phase retardation pattern (right) on a Spatial Light Modulator.}
\end{figure}

A phase-only SLM consists of an array of liquid-crystal pixels which can be individually oriented.  These can be used to manipulate the profile of a laser beam in the far-field by imparting a controlled, spatially-varying retardation on the light.  The far-field pattern can also be realized by focusing the phase-modulated laser beam using a lens.

For a given target intensity distribution in the focal plane of the lens, we use the Mixed-Region Amplitude-Freedom (MRAF) algorithm  \cite{Pasienski_08} to calculate the optimal phase pattern in the SLM plane. This algorithm performs an optimization procedure on a guess phase pattern, by iteratively transforming the light field between the SLM plane and the focal plane of the lens (i.e. the output plane) by means of Fast Fourier Transforms (FFTs).  After each FFT, the desired intensity of the light field is enforced on both planes. In particular, the MRAF algorithm divides the output plane into two regions: a signal region in which we restrict the intensity to match our target pattern, and a noise region in which the intensity is unconstrained.  This separation allows for increased accuracy in the signal region. Phase freedom in the output plane is permitted, while the intensity profile in the input plane is that of the laser beam illuminating the SLM. The algorithm allows accurate and fast (a few hundred iterations) calculations of phase patterns to generate almost any arbitrary light pattern in the signal region of the output plane. An example of a phase pattern that generates a  power-law intensity distribution of order $\alpha=2$ is shown in Fig. \ref{fig:Output}.

\begin{figure}%[t]
\includegraphics[width=0.45\textwidth]{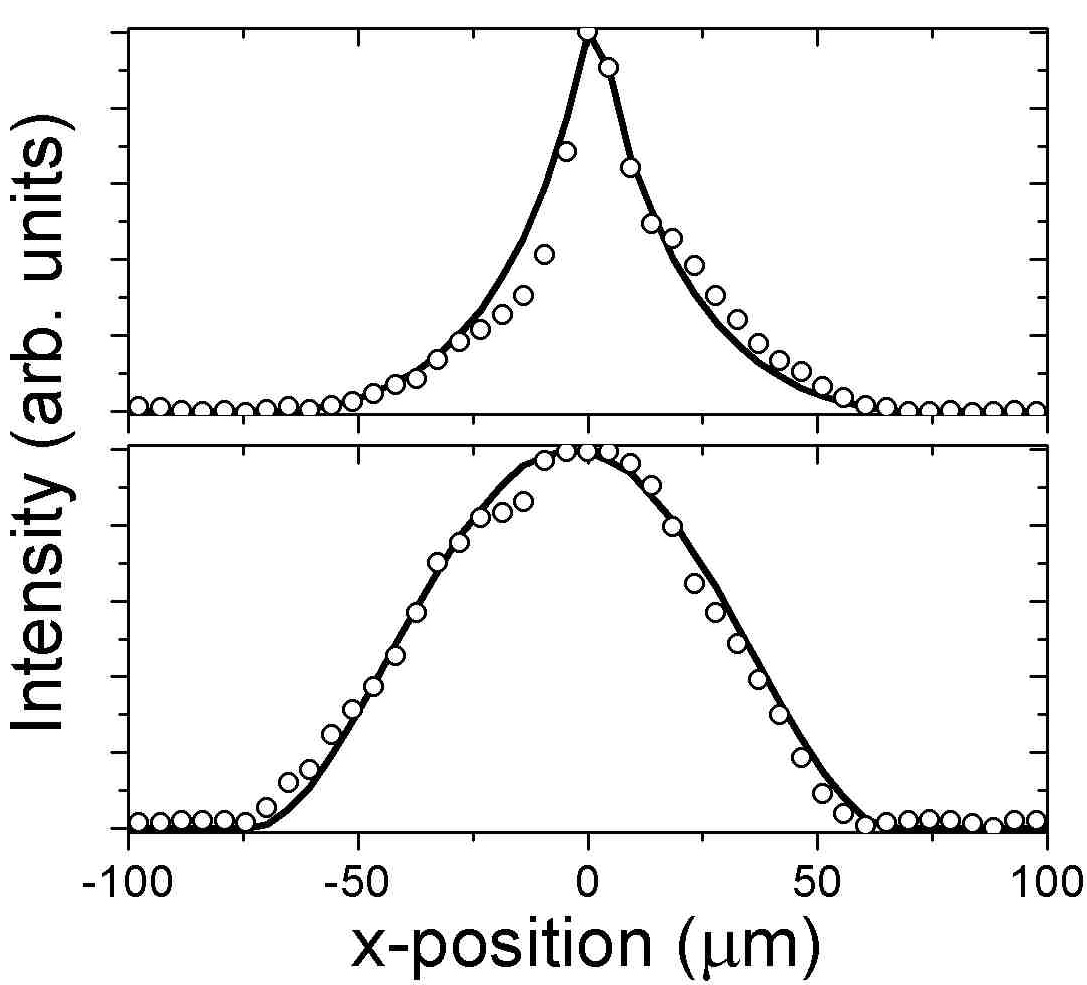}
\caption{\label{fig:BeamOrders}Profiles of power-law intensity distributions with $\alpha = 0.5$ (top) and $\alpha = 2$ (bottom). The measured curves (open circles) are extracted from images such as the one shown in Fig. \ref{fig:Output}, while the predicted curves (solid lines) are determined by the MRAF algorithm.}
\end{figure}

The optimal retardation pattern calculated using the above method is then applied to an SLM (BNS P-256), which contains $256 \times 256$ $24\mu$m nematic liquid crystal pixels. The SLM is illuminated with $1060~$nm light generated by a diode laser, and the diffracted beam, focused by an $f=50~$mm achromatic doublet lens, is detected by a CCD camera. The MRAF-optimised phase pattern does not always produce an accurate intensity pattern when applied to a physical device, due to imperfect device response and aberrations in the optical system \cite{Pasienski_08, Bruce_11}. However for the case of simple power-law intensity distributions, we have found that the output is smooth and accurate. In particular we have generated power laws with $\alpha$ ranging between $2$ and $0.5$. Fig. \ref{fig:Output} shows the light pattern in the output plane as detected by the CCD camera, while Fig. \ref{fig:BeamOrders} shows the intensity profiles of $\alpha=2$ (parabolic) and $\alpha=0.5$ distributions. At fixed order, we have also varied the radius $r_{0}$ between $27~\mu$m and $103~\mu$m. The size of the signal region is chosen so that the uncontrolled intensity in the noise region is well separated from the power-law pattern. This minimises the effect of the noise region on atoms loaded in the power-law trap. 

%\begin{figure}[ht!]

\begin{figure}[t]
\includegraphics[width=0.45\textwidth]{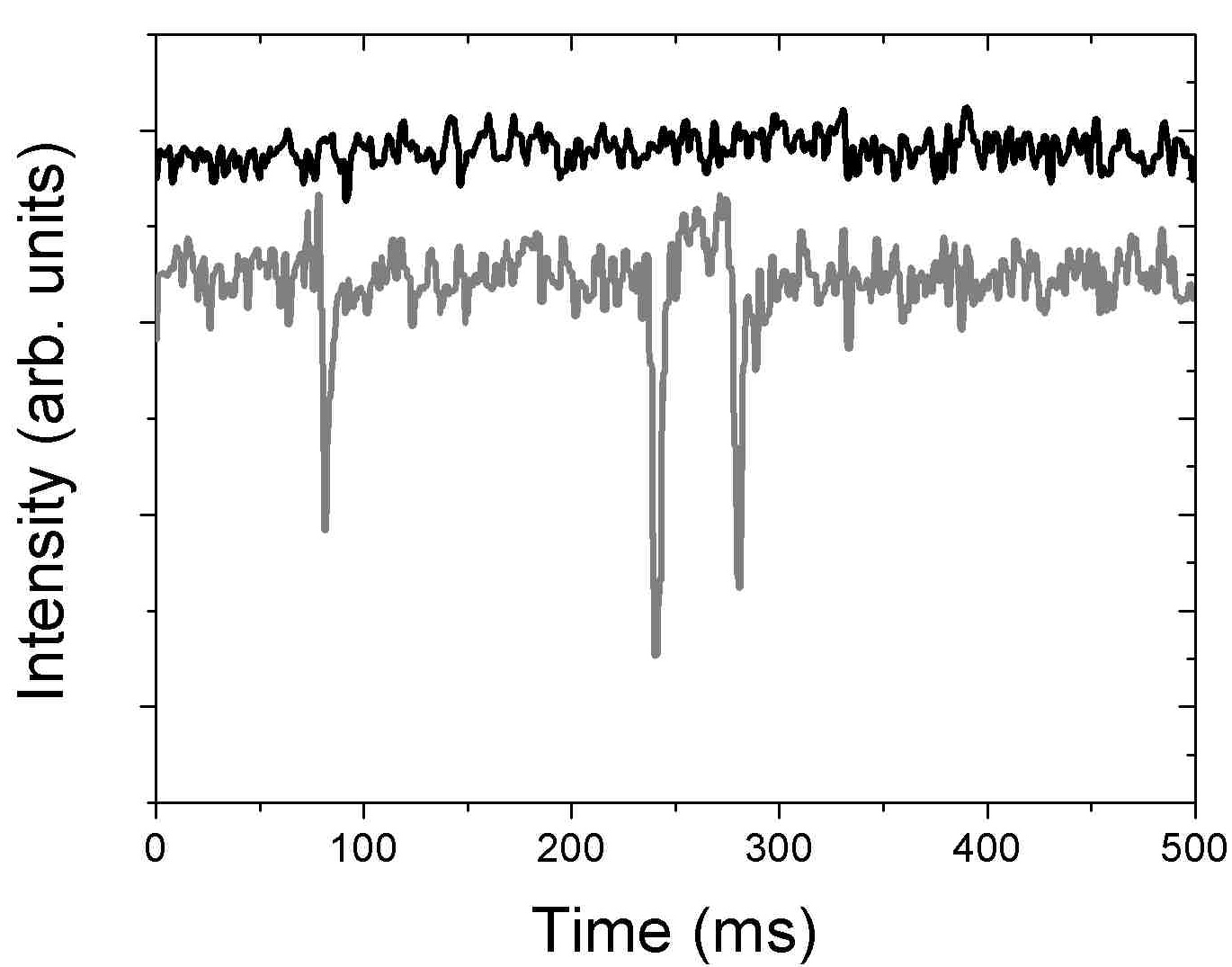}
\caption{\label{fig:flicker}Substantial flicker can occur as the SLM switches between patterns (lower trace).  By minimising the level of phase change per SLM pixel, this flicker can be removed (upper trace).}
\end{figure}

By applying a sequence of phase patterns to the SLM, dynamic optical traps can be generated. It has been noted previously \cite{Boyer_04} that dynamic light patterns generated by ferroelectric liquid crystal SLMs can be subject to substantial intensity flicker due to the changes in the state of individual pixels. We find that this problem also exists in our nematic liquid crystal SLM, but can be solved by careful implementation of the MRAF algorithm.  To achieve dynamic power-law optical traps, we apply a sequence of phase patterns producing power laws withr $\alpha$ going from $0.5$ to $2$ in steps of $0.1$ at $25~$Hz.  In the lower trace in Fig. \ref{fig:flicker}, intensity flicker occurs when the order changes from $0.5$ to $0.6$ and also to and from order $1$. This is because the initial guess phase used as the input to the MRAF algorithm for orders $0.5$ and $1$ happens to be different from the one used for all the other orders. Since the MRAF-optimised phase pattern strongly depends on the initial guess, a significant change is incurred in moving from one of these orders to any of the others. In particular $99.7\%$ of the SLM pixels change the phase shift they impart to the light, by $0.7\pi$ on average. To overcome this, the retardation patterns are recalculated using the same guess phase for all orders. This results in a sequence in which $70\%$ of pixels change their phase shift at each step, but only by $0.025\pi$ on average. As shown in the upper trace of Fig. \ref{fig:flicker}, we can no longer measure any flicker between consecutive patterns.

%\begin{figure}[htb!]

To ensure that the two-dimensional intensity patterns generated by the SLM are suitable for three-dimensional optical trapping of cold atoms, we verified that the beam maintains its power-law profile for a propagation of at least $300~\mu$m away from the focal plane of the lens. This distance is greater than the width of all the power-law distributions we realised, hence two perpendicular power-law beams, each tailored by an SLM, can provide three-dimensional confinement as discussed later in this paper. Alternatively, a light sheet can be used to add confinement along the axis of the SLM beam.

\section{Adiabaticity Conditions} \label{sect:adiabaticity}

In the previous section, we showed that the size and order of a power-law trap can be dynamically varied by refreshing the phase pattern on the SLM. Moreover, the trap depth can be controlled by varying the optical power illuminating the SLM. This gives full flexibility in designing sequences. How fast a sequence can then be implemented is limited by the requirement that changes in the trapping potential must be slow compared to the thermalisation time, which is determined by the elastic collision rate. Additionally, changes must be slow with respect to the motion of the atoms in the trap. In the case of a harmonic trap characterised by a frequency $\omega$, any compression or expansion of the trap must satisfy $\frac{d \omega/\omega}{ d t} << \omega$ \cite{Ketterle_99}.

Similarly, in our generic power-law trap such an adiabaticity condition can be formulated for each of the three parameters (radius, depth and order), given that they can be varied independently. The condition therefore is that the relative change of a parameter per unit time be much less than the characteristic oscillation frequency of an atom in the trap. For the traps considered in this work, this is of the order of $\sqrt{A/m}/r_{0}$ where $m$ is the atomic mass \cite{Landau_93}. Hence the adiabaticity conditions can be written as:

\begin{equation}
\frac{\Delta k/k}{\Delta t}\ll \frac{1}{r_{0}}\sqrt{\frac{A}{m}}, \label{adiab}
\end{equation}

\noindent where $k=A,r_{0},\alpha$. The SLM can vary $\alpha$ and $r_{0}$ in small but discrete steps, and it takes $\Delta t \approx 10~$ms to change from one pattern to the next, as estimated by the duration of the flicker shown in Fig. \ref{fig:flicker}. For the transformations considered below, $\Delta \alpha = 0.1$, $\Delta r_{0} \leq 3~\mu$m, and the characteristic oscillation frequencies are several KHz. These parameters ensure that Eq. (\ref{adiab}) is satisfied for $k=r_{0},\alpha$. The trap depth on the other hand can be varied continuously, hence $\Delta A$ can be taken as the total change over a sequence, and $\Delta t$ as the total duration of that sequence. We expect the evaporative cooling sequence and the adiabatic transformation described below to last about one second, for which Eq. (\ref{adiab}) is satisfied.

\section{Evaporative cooling sequence} \label{sect:evap}

\begin{figure*}%[htb!]
\includegraphics[width=0.93\textwidth]{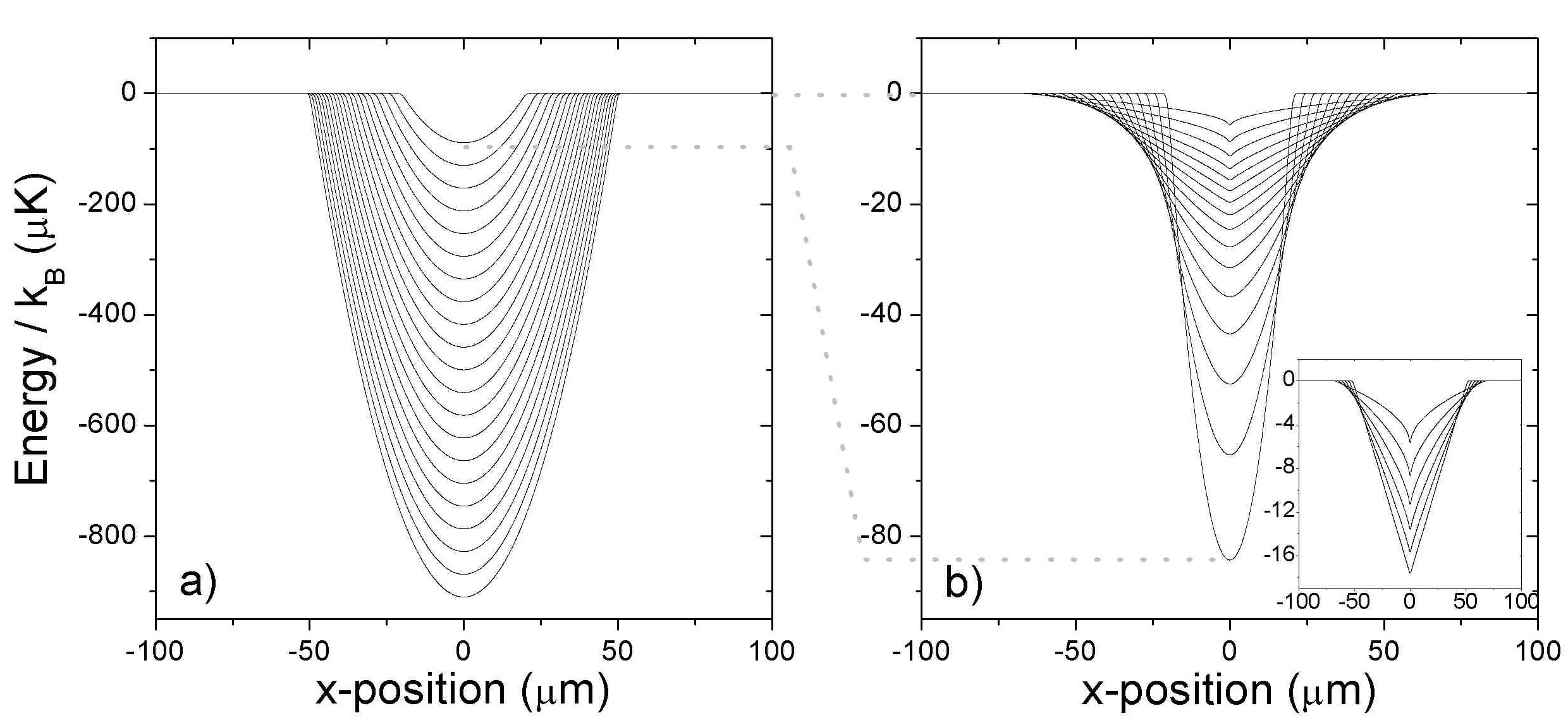}
\caption{\label{fig:Profile}The evolution of the trap profile for creating a Bose--Einstein condensate.  (a) A power-law trap of order 2 is decreased in depth and width such that atoms are evaporatively cooled at constant elastic collision rate. (b) Following evaporation, the trap is adiabatically transformed from order 2 to order 0.5 to reach the critical temperature at the end of the sequence. The inset shows in greater detail the final stages of the adiabatic transformation.}
\end{figure*}

In the remainder of the paper, we assume that two power law-shaped laser beams are crossed to form a three-dimensional trap. We assume for simplicity that this trap is spherically symmetric (which can be realized by modifying the aspect ratio of the light distributions shown in Sec. \ref{sect:generate}), but our results are also valid in the case of asymmetric traps as long as the trap order is the same in all directions. 

As for any all-optical scheme, our approach is suitable for all internal states, atomic mixtures and molecules. In the following we use $^{87}$Rb as an example. Laser-cooled atoms from a magneto-optical trap can be captured into a broad (radius $r_{0} = 50~\mu$m), $910 ~\mu$K deep parabolic ($\alpha=2$) optical trap. This is the deepest trap shown in Fig. \ref{fig:Profile}a. Following Eqs. (\ref{eqn:I}) and  (\ref{eqn:trappingpot}), an optical power of $14~$W in each beam is needed for such a trap at a wavelength of $1060~$nm. 

From experiments in which similar conditions have been realized in practice \cite{Jacob_11}, we take as our starting point $N= 3 \times 10^{5}$ atoms with a temperature $T = 91~ \mu$K, resulting in a phase-space density  $D = 10^{-4}$ and an elastic collision rate $\kappa \sim 2000 ~$s$^{-1}$. This elastic collision rate is sufficiently high that a first stage of adiabatic compression is not necessary. However this could be included (by reducing the trap radius and increasing the depth at constant power) if the starting conditions required it, hence evaporation can be optimized for a broad range of experimental parameters. We note moreover that the chosen initial atom number is conservative, and that increasing the trap size results in more atoms being captured \cite{Jacob_11}, ultimately leading to a larger condensate. 

A possible evaporation sequence is shown in Fig. \ref{fig:Profile}a. The trap depth is gradually lowered to force evaporation, while the trap radius is adjusted to maintain optimal conditions. In particular, we choose to keep the elastic collision rate constant, as opposed to selecting a sequence of runaway evaporation. Given our starting conditions, this ensures that the atomic number density does not increase to the point where three-body recombination losses become significant. The trap order is kept at 2 because smaller values would result in decreased efficiency of the evaporation process \cite{Ketterle_96}. The efficiency is defined as:

\begin{equation}
\gamma = - \frac{ d \ln D }{ d \ln N }, 
\end{equation}

\noindent which is equivalent to $N \propto 1/D^{\frac{1}{\gamma}} \propto T^{\frac{3}{2\gamma}} V^{\frac{1}{\gamma}}/N^{\frac{1}{\gamma}}$, where $V$ is the effective volume occupied by the atoms. This relation can be used to express the dependence of $N$ on $V$ and $T$, which is then substituted in the condition for the elastic collision rate $\kappa \propto T^{1/2}N/V=$ constant. Next we express $T$ and $V$ in terms of the trap depth and radius respectively: $T \propto A$ and $V \propto r_{0}^{3}$, assuming the truncation parameter $\eta = A/(k_{B}T)$ is kept fixed throughout evaporation. The condition for constant $\kappa$ finally becomes:

\begin{equation}
\frac{A^{\frac{\gamma}{2} + 2}}{ r_{0}^{3 \gamma}}=\text{constant},
\end{equation}

\noindent which is used to plot the intermediate steps of the evaporation sequence in Fig. \ref{fig:Profile}a. For this we assume $\gamma = 3$ and $\eta=10$ (i.e. evaporation near stagnation), as typical for evaporation in optical traps \cite{OHara_01,Kinoshita_05,Clement_09,Jacob_11}. Similarly to these experiments, we expect evaporation to work on timescales of seconds. Hence the SLM needs to be refreshed at a rate less than $20~$Hz to produce the intermediate steps. Given that this is much smaller than the characteristic trap frequency, we do not expect significant parametric heating \cite{Savard_97} from residual intensity flicker at the refresh rate. 

At the end of the evaporation sequence, the final trap radius and depth are $20~\mu$m and $84~ \mu$K  respectively, at which point $D=0.011$ and $N=6 \times 10^4$ atoms remain in the trap. The adiabatic transformation described in the next section is then used to bring the sample to degeneracy.

\section{Adiabatic Transformations} \label{sect:transform}

As shown in \cite{Pinkse_97}, the phase-space density of a collisional gas can be increased adiabatically and reversibly, without loss of atoms, by reducing the trap order. This is different from compressing or expanding a trap while keeping the same order, in which case the temperature and the number density change so as to keep the phase-space density constant. In the following, we consider an adiabatic transformation in which the trap order goes from $\alpha=2$ to $0.5$, starting from the parabolic trap at the end of the evaporation sequence (i.e. the smallest trap in Fig. \ref{fig:Profile}a). We define our transformation such that at $\alpha = 0.5$ the atoms reach the critical temperature for Bose-Einstein condensation, i.e. $T=T_{c}$ and $D=2.612$.

%\begin{figure}[htb!]
\begin{figure}%[t]
\includegraphics[width=0.45\textwidth]{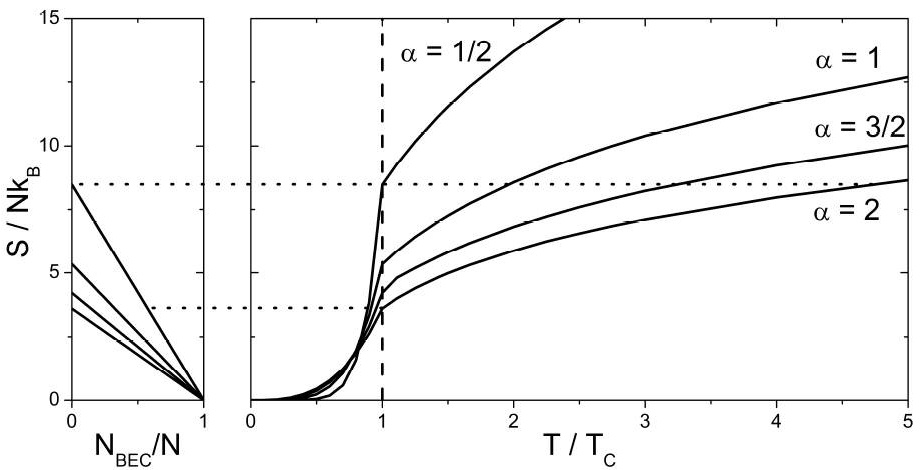}
\caption{\label{fig:Entropy}The entropy of a Bose gas in a power-law trap of order $\alpha$ as a function of (left) condensate fraction and (right) temperature above and below $T_{c}$.  The two dotted lines represent transformations at constant entropy. By transforming from a power-law trap of order 2 containing a gas at $T/T_{c}=4.7$ to a power-law trap of order 0.5, the gas reaches the critical temperature (upper dotted line). Starting from $T=T_{c}$ in the $\alpha=2$ trap, an adiabatic transformation to $\alpha=0.5$ leads to a condensate fraction of about 0.6 (lower dotted line).}
\end{figure}

For such a transformation to be reversible, it must be slow compared to the thermalization time. We estimate that the elastic collision rate during our sequence does not drop significantly from the value of $2000 ~$s$^{-1}$ during evaporation, so a sequence lasting about one second (as suggested in Sec. \ref{sect:adiabaticity}) satisfies this requirement. We can therefore impose that the entropy $S$ is constant during the transformation, with $S$ defined by \cite{Pinkse_97}:

\begin{equation}
\frac{S}{N k_{B}} = \label{Sabove}
\begin{cases}
\left(\frac{5}{2}+\frac{3}{\alpha}\right)\frac{g_{\frac{5}{2}+\frac{3}{\alpha}}\left(z\right)}{g_{\frac{3}{2}+\frac{3}{\alpha}}\left(z\right)}-\ln\left(z\right) \ \ \  & T\geq T_{C}, \\
\left(\frac{5}{2}+\frac{3}{\alpha}\right)\frac{g_{\frac{5}{2}+\frac{3}{\alpha}}\left(1\right)}{g_{\frac{3}{2}+\frac{3}{\alpha}}\left(1\right)}\left(\frac{T}{T_{C}}\right)^{\frac{3}{2}+\frac{3}{\alpha}} & T\leq T_{C}, 
\end{cases}
\end{equation}

\noindent where $z = \exp\left(\mu/k_{B}T\right)$ is the fugacity with $\mu$ the chemical potential, and $g_{\kappa}\left(z\right) = \sum^{\infty}_{j=1}z^{j}j^{-\kappa}$. Below $T_{c}$, $\mu=0$ and therefore $z=1$. Above $T_{c}$, we find $z$ by numerically solving \footnote{Eq. (\ref{eqn:findz}) is obtained from Eq. (4) in \cite{Pinkse_97} at and above $T_{c}$.}:

\begin{equation}
g_{\frac{3}{2}+\frac{3}{\alpha}}\left(z\right) = g_{\frac{3}{2}+\frac{3}{\alpha}}\left(1\right) \left(\frac{T_{c}}{T}\right)^{\frac{3}{2}+\frac{3}{\alpha}}. \label{eqn:findz}
\end{equation}

This results in the plots of entropy per particle versus $T/T_{c}$ shown in Fig. \ref{fig:Entropy} for several trap orders. For the conservation of entropy, the value $S/Nk_{B}=8.47$ at the critical temperature for $\alpha=0.5$ (see upper dotted line in Fig. \ref{fig:Entropy}) must also  be the entropy per particle at the beginning of the adiabatic transformation, i.e. for $\alpha=2$. This corresponds to $T/T_{c}=4.7$ and $D_{1}=0.011$ in the parabolic trap, as achieved at the end of the evaporation. From the conservation of entropy and Eq. (\ref{eqn:findz}) we can then determine $z(\alpha)$ and $\frac{T}{T_{c}}(\alpha)$ for the intermediate steps of the adiabatic sequence. 

\begin{figure}%[t]
\includegraphics[width=0.44\textwidth]{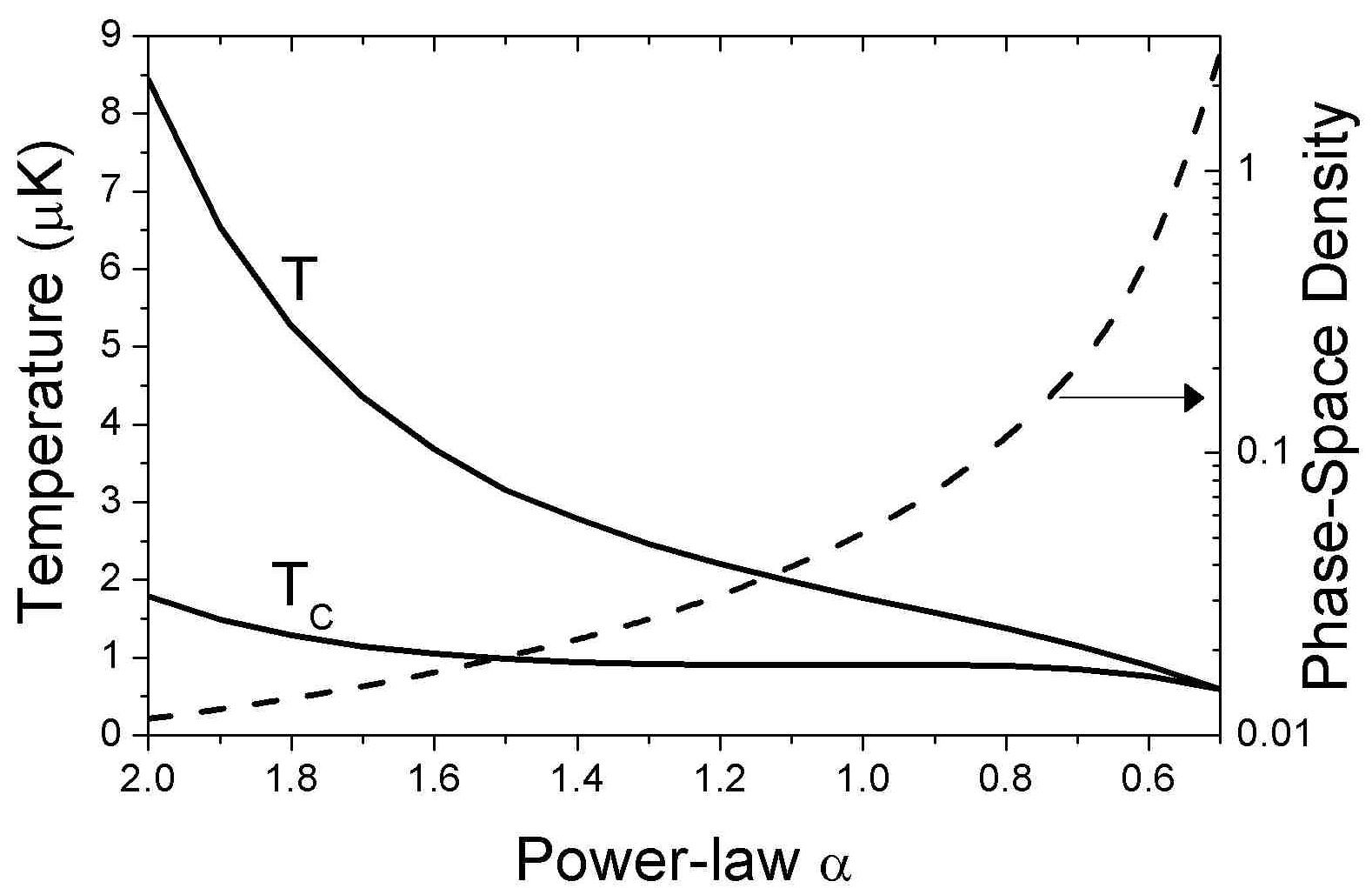}
\caption{\label{fig:Temperature}The temperature and critical temperature decrease by changing the power-law order $\alpha$, until they are equal.  The phase-space density of the gas increases by about a factor 230 to 2.612.}
\end{figure}

To calculate the trap depth $A(\alpha)$ and size $r_{0}(\alpha)$ required for this sequence, we start by imposing $A(\alpha)=10 k_{B} T(\alpha)$ to avoid further evaporation during the transformation. Secondly, the end point of the sequence is determined by imposing a peak density $n=1.8 \times 10^{14}~$cm$^{-3}$, which gives a moderate rate of atom loss due to three-body recombination of $1$s$^{-1}$ \cite{Marte_02}. We then use the critical condition $n \lambda_{dB}^3 =2.612$ (with the thermal de Broglie wavelength $\lambda_{dB} = h/\sqrt{2 \pi m k_{B} T}$) to calculate $T_{c}= 582~$nK in the $\alpha = 0.5$ trap.  To find the trap size $r_{0}(0.5)$, we consider the expression for $T_{c}$ in a generic power-law trap \cite{Dalfovo_99}:

\begin{equation}
k_{B} T_{c} = \left[\frac{N \hbar^3}{\left(2m\right)^{ \frac{3}{2}}} \frac{6\sqrt{\pi}A^{\frac{3}{\alpha}}}{r_{0}^{3} \Gamma\left(1+\frac{3}{\alpha}\right)\zeta\left(\frac{3}{2}+\frac{3}{\alpha}\right)}\right]^{\frac{1}{\frac{3}{2}+\frac{3}{\alpha}}}, \label{Tc}
\end{equation}

\noindent and we solve it for $r_{0}$  to find $r_{0}(0.5)=59~\mu$m. Having fixed the trap parameters at the start and the end of the sequence, we assume for simplicity a linear interpolation for $r_{0}(\alpha)$. We use Eq. (\ref{Tc}), in conjunction with $\frac{T}{T_{c}}(\alpha)$ from entropy conservation, to determine $T(\alpha)$ and $T_{c}(\alpha)$ individually, as shown in  Fig. \ref{fig:Temperature}.  The resulting trap profiles are shown in Fig. \ref{fig:Profile}b.

%\begin{figure}[htb!]

The optical power needed for the final trap ($\alpha=0.5$) is $36~$mW per beam, which is $0.26\%$ of the power at the start of evaporation. Such a large dynamic range can be achieved by varying the power illuminating the SLM, e.g. with a motorized rotating waveplate followed by a Glan-Taylor polarizer, and by changing the power emitted by the laser \cite{Jacob_11}. Towards the end of the adiabatic transformation, the trap is so shallow that the effect of gravity cannot be neglected. It is possible to compensate for it either with a magnetic field gradient, or by designing ``tilted'' intensity patterns. 

We see from Fig. \ref{fig:Temperature} that the adiabatic transformation achieves a significant gain in phase-space density. Intuitively, the gas undergoes an expansion and its temperature decreases. However the change in trap order causes an increase in \emph{peak} density which is especially pronounced at the end of the sequence. We estimate that only $5\%$ of atoms are lost due to three-body recombination for a sequence that lasts $1$s, which justifies our assumption of constant atom number throughout the adiabatic transformation. In conclusion, about $20\%$ of the laser-cooled atoms loaded into the initial optical trap remain when the sequence terminates at the BEC transition, which is higher than in any other all-optical technique implemented so far.  

If evaporation is continued to the BEC transition at the same efficiency assumed in Sec. \ref{sect:evap}, $10^4$ atoms remain in the trap - which is only $3\%$ of the initial atom number. In this case however the adiabatic change of trap order (see lower dotted line in Fig. \ref{fig:Entropy}) provides a new method to cross the BEC transition in a reversible way, similarly to the dimple technique \cite{Stamper-Kurn_98}. From the conservation of entropy, we find that varying $\alpha$ from $2$ to $0.5$ leads to $T/T_{c}\simeq 0.89$. A condensate fraction $N_{BEC} / N \simeq 0.6$ is found using $N_{BEC}/N = 1 - \left(T/T_{c}\right)^{3/2+3/\alpha}$ \cite{Pinkse_97}, as shown in the left side of Fig. \ref{fig:Entropy}, hence $N_{BEC}\simeq 6000$ at the end of the adiabatic transformation.  Our sequence could then be combined with \emph{in-situ} imaging \cite{Andrews_96} to provide a reversible method to investigate the BEC transition.

\section{Conclusions}

We have used an SLM to holographically generate power-law intensity patterns of different orders and sizes, and we have shown how a sequence of these can be used as a dynamic optical trap for fast and efficient production of Bose-Einstein condensates. Starting from realistic assumptions, we have calculated the trap parameters throughout the sequence. We have also presented the adiabaticity criteria for a generic transformation that goes beyond the standard case of the compression and expansion in a harmonic trap. A future step will be the study of evaporative cooling and adiabatic transformations in more complex trap geometries. 		
		
\begin{acknowledgments}
The authors would like to thank Steve Lister, James Mayoh and Tiffany Harte for useful discussions.  This work was supported by the UK EPSRC, and GS acknowledges support from a SUPA Advanced Fellowship.
\end{acknowledgments}

\bibliography{PowerLawBib}

\end{document}